\documentclass[aps,pra,twocolumn,amsmath,amssymb,nofootinbib,superscriptaddress]{revtex4}

\newcommand{\ket}[1]{|#1\rangle}

\usepackage[pdftex]{graphicx}
\usepackage{mathrsfs}
\usepackage{color}
\usepackage[colorlinks]{hyperref}

\begin{document}

\bibliographystyle{apsrev}

%
%

\title{Spontaneous parametric down-conversion photon sources are scalable in the asymptotic limit for boson-sampling}

%
%

\author{Keith R. Motes}
\affiliation{Centre for Engineered Quantum Systems, Department of Physics \& Astronomy, Macquarie University, Sydney NSW 2113, Australia}

\author{Jonathan P. Dowling}
\affiliation{Hearne Institute for Theoretical Physics and Department of Physics \& Astronomy, Louisiana State University, Baton Rouge, LA 70803}
\affiliation{Computational Science Research Center, Beijing 100084, China}

\author{Peter P. Rohde}
\email[]{dr.rohde@gmail.com}
\homepage{http://www.peterrohde.org}
\affiliation{Centre for Engineered Quantum Systems, Department of Physics \& Astronomy, Macquarie University, Sydney NSW 2113, Australia}

\date{\today}

\frenchspacing

%
%

\begin{abstract}
Boson-sampling has emerged as a promising avenue towards post-classical optical quantum computation, and numerous elementary demonstrations have recently been performed. Spontaneous parametric down-conversion (SPDC) is the mainstay for single-photon state preparation, the technique employed in most optical quantum information processing implementations to-date. Here we present a simple architecture for boson-sampling based on multiplexed SPDC sources and demonstrate that the architecture is limited only by the post-selection detection efficiency assuming that other errors, such as spectral impurity, dark counts, and interferometric instability are negligible. For any given number of input photons, there exists a minimum detector efficiency that allows post selection. If this efficiency is achieved, photon-number errors in the SPDC sources are sufficiently low as to guarantee correct boson-sampling most of the time. In this scheme the required detector efficiency must increase exponentially in the photon number. Thus, we show that idealised SPDC sources will not present a bottleneck for future boson-sampling implementations. Rather, photodetection efficiency is the limiting factor and thus future implementations may continue to employ SPDC sources.
\end{abstract}

\maketitle

%
%

\section{Introduction}

Linear optics quantum computing (LOQC) \cite{bib:KLM01, bib:Kok07, bib:KokLovett11} is a promising route towards scalable universal quantum computing \cite{bib:NielsenChuang00}. The first architecture, presented by Knill, Laflamme \& Milburn (KLM) \cite{bib:KLM01}, demonstrated that scalable quantum computation is possible using only single-photon sources, photodetection, quantum memory and fast-feedforward. However, the physical resource requirements are daunting, and large-scale LOQC appears distant. Since the advent of KLM, numerous simplifications have been suggested, significantly reducing physical resource requirements \cite{bib:Nielsen04, bib:BrowneRudolph05}, but nonetheless require technologies such as quantum memory and fast-feedforward that are not presently available.

Recently, Aaronson \& Arkiphov \cite{bib:aaronson2011computational} presented an alternate linear optical scheme, known as boson-sampling. This scheme is believed to implement a classically hard algorithm for a specific task, but will likely not be universal for quantum computation. In this model, only single-photon state preparation, passive linear optics (beamsplitters and phase-shifters), and photodetection are required, doing away with the more challenging requirements of fast-feedforward and quantum memory. The system's Hilbert space scales exponentially with the physical resources. Gard \emph{et al.} present an elementary argument from a quantum optics perspective as to why boson-sampling scales exponentially \cite{bib:GardCrossDowling}. 

The technology to implement boson-sampling is, for the larger part, available today, making boson-sampling an attractive route towards a type of non-universal optical quantum information processing. Recently, numerous experimental groups have begun implementing elementary demonstrations of boson-sampling using only a few photons \cite{bib:metcalf12, bib:Broome20122012, bib:Spring2, bib:Crespi3, bib:Tillmann4}.

While boson-sampling is one of the first non-trivial computational problems solvable with a linear optical interferometer using Fock-state inputs, it is likely not the last. Boson-sampling is a completely new quantum computational scheme that has yet to be fully explored and understood. Fully understanding boson-sampling may present us with new computational problems not accessible by classical computers. Furthermore, the exponentially large Hilbert space and computational complexity associated with such interferometers will likely lead to further breakthroughs in the closely related fields of quantum optical metrology, imaging, and sensing \cite{bib:lee2002quantum, bib:dowling2008quantum}.

In this paper we show that large scale boson-sampling can be implemented provided that detection efficiencies, which must increase exponentially with photon number, are sufficient to guarantee post-selection with high probability. Increasing input photon number will thus yield a larger required detection efficiency. 

Spontaneous parametric down-conversion (SPDC) has become the mainstay for single-photon state preparation, is widely used in optical quantum information processing, and was employed in all of the recent experimental boson-sampling implementations. A pressing question for future larger-scale implementations is scalability. Scalability in this context refers to increasing the input photon number into the boson-sampling device provided that the error in the single photon photo-detectors, which scales exponentially with input photon number, is sufficiently low to ensure successful implementation of boson-sampling most of the time. That is, what are the limitations and requirements on physical resources to implement a scalable device? In particular, will SPDC sources suffice, or will we have to transition to other photon source technologies? The issue of scalability of SPDC sources in the context of boson-sampling was recently discussed by Lund \emph{et al.} \cite{bib:lund2013boson}.

We consider a general architecture for the experimental implementation of boson-sampling, where multiplexed SPDC sources are employed for state preparation. We show that in such an architecture the device is limited only by the post-selection probability. In other words, the architecture is scalable provided that detector efficiencies are sufficiently high to enable post-selected computation. In this regime, the quality of current SPDC states is sufficient to enable large-scale boson-sampling. Thus, it is photodetection, not SPDC sources, that provide the bottleneck to larger-scale demonstrations.

%
%

\section{The boson-sampling model}


We begin by preparing an $m$-mode state, in which the first $n$ modes are initialised with single-photon Fock states and the remainder in the vacuum state,
\begin{eqnarray} \label{eq:input_state}
\ket{\psi}_\mathrm{in} &=& \ket{1_1,\dots,1_n,0_{n+1},\dots,0_m} \nonumber \\
&=& \hat{a}_1^\dag \dots \hat{a}_n^\dag \ket{0_1,\dots,0_m},
\end{eqnarray}
where $\hat{a}_i^\dag$ is the photon creation operator in the $i$th mode, and $m=O(n^2)$. This state is manipulated via a passive linear optics network which implements a unitary map on the photon creation operators,
\begin{equation}
\hat{a}_i^\dag \to \sum_{j=1}^m U_{ij} \hat{a}_j^\dag,
\end{equation}
where $U$ is an $m\times m$ unitary matrix. It was shown by Reck \emph{et al.} \cite{bib:Reck94} that any $U$ can be efficiently constructed using $O(m^2)$ linear optics elements.

In an occupation-number representation, the output state is of the form,
\begin{equation}
\ket{\psi}_\mathrm{out} = \sum_S \gamma_S \ket{n_1^{(S)},\dots,n_m^{(S)}},
\end{equation}
where $S$ are the different photon number configurations, the number of which grows exponentially with the number of photons, as $|S|=\binom{n+m-1}{n}$, and $n_i^{(S)}$ is the number of photons in mode $i$ associated with configuration $S$.

\begin{figure}[!htb]
\includegraphics[width=0.4\columnwidth]{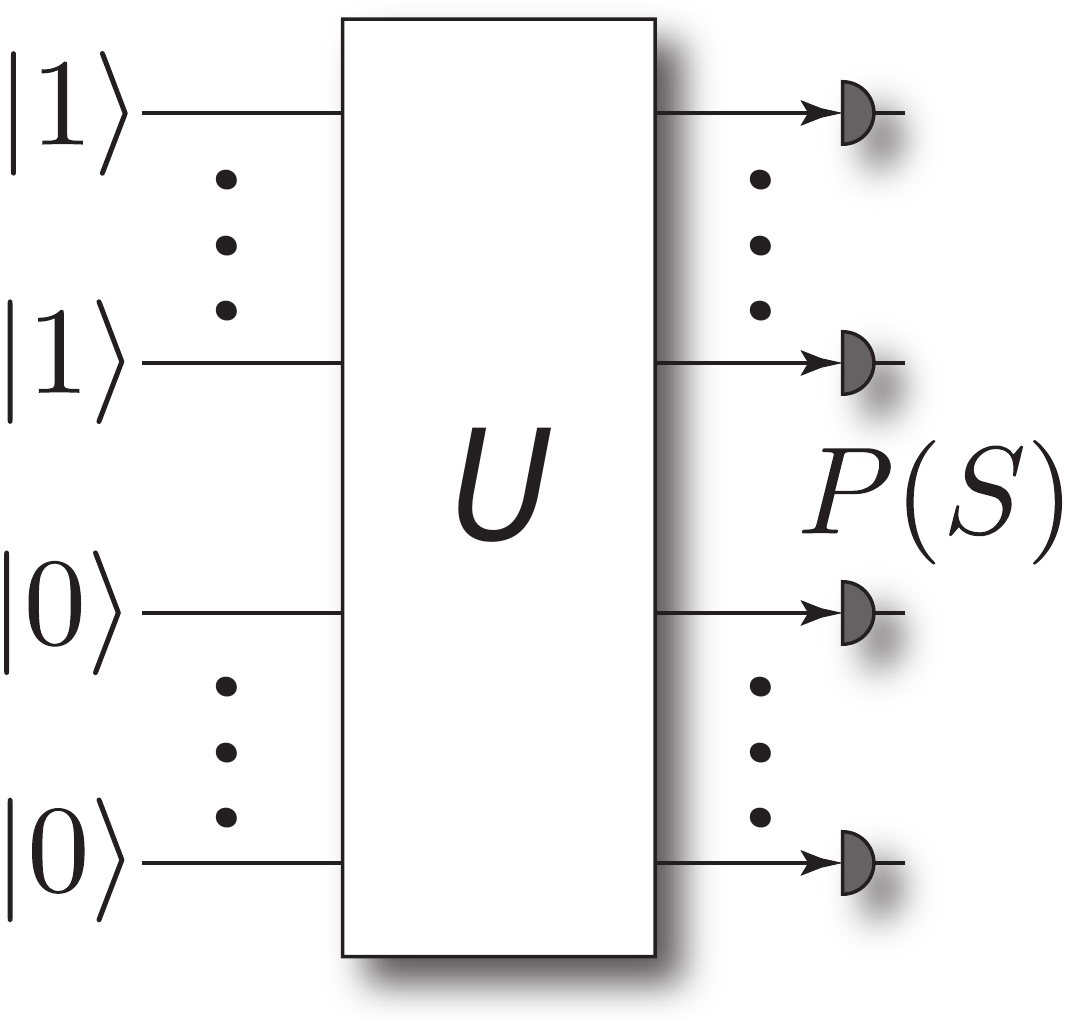}
\caption{The boson-sampling model. The input state is prepared, comprising a number of single-photon Fock states and vacuum states. The input state passes through a passive linear optics network $U$ comprising beamsplitters and phase-shifters. Finally, the experiment is repeated many times, and each time the output photon number statistics are sampled from $P(S)$ via coincidence number-resolving photodetection.} \label{fig:model}
\end{figure}

Finally, we perform number-resolved photodetection \cite{bib:rosenberg2005noise} on the output distribution, obtaining a sample from the distribution $P(S)=|\gamma_S|^2$. The experiment is repeated many times, building up statistics of the output distribution. It was shown by Aaronson \& Arkhipov \cite{bib:aaronson2011computational} that this sampling problem likely cannot be efficiently simulated classically. The intuitive explanation for this supposed classical hardness is that each of the amplitudes $\gamma_S$ is proportional to an $n\times n$ matrix permanent. Permanents are believed to be classically hard to calculate, residing in the complexity class \mbox{\textbf{$\#$P}-complete}, the class of polynomial time counting problems. The boson-sampling model is illustrated in Fig.~\ref{fig:model}.

Boson-sampling is not believed to be capable of efficiently simulating full quantum computation. Nonetheless, it is a relatively simple scheme that can likely rival classical computers for certain tasks, thus it is an attractive post-classical quantum computation scheme. It was shown by Rohde \& Ralph that boson-sampling may implement a computationally hard algorithm even in the presence of high levels of loss \cite{bib:RohdeRalphErrBS} and mode-mismatch \cite{bib:RohdeLowFid12}, although formal hardness proofs are still lacking.

%
%

\section{Experimental architecture for boson-sampling}

Given that SPDC is the most widely used and readily accessible source for single-photon state preparation, we will present a simple architecture for boson-sampling based on SPDC sources. In an ideal boson-sampling implementation one would employ deterministic photon sources that produce exactly one photon on demand. SPDC sources, on the other hand, coherently prepare photon pairs in two modes with a correlated Poisson probability distribution. By measuring one of the modes and post-selecting upon detecting one photon in that mode, a single photon is guaranteed to appear in the other mode. This method provides us with a probabilistic, but heralded single-photon source. It is critical that each photon is heralded to ensure a pure set of Fock-state inputs. 

Specifically, the two-mode state prepared by a SPDC is of the form,
\begin{equation}
\ket{\psi}^\mathrm{SPDC} = \sum_s \lambda_s \ket{s,s},
\end{equation}
and the photon number probability distribution is given by \cite{bib:GerryKnightBook},
\begin{equation} \label{eq:PDC_dist}
P^\mathrm{SPDC}(s) = |\lambda_{s}|^2 = \frac{\mathrm{tanh}^{2s}r}{\mathrm{cosh}^{2}r},
\end{equation}
where $s$ is the photon number (per mode) and $r$ is the squeezing parameter. Thus, the SPDC source most often emits the vacuum state, and sometimes higher-order pairs with exponentially decreasing probability. For small squeezing parameters the higher-order terms can be made small, yielding a heralded source that produces single pairs with high probability.

To herald a single photon, we detect one arm of a single SPDC source using an inefficient number-resolving photodetector. Such a detector can be characterised by the conditional probability of detecting $t$ photons given that $s$ photons were present. For a simple inefficient detector this is given by,
\begin{equation} \label{eq:det_mn}
P_\mathrm{D}(t|s) = \binom{s}{t} \eta^t (1-\eta)^{s-t},
\end{equation}
where $\eta$ is the detection efficiency. Thus, in the presence of loss, the detector exhibits ambiguity in the measured photon number, sometimes detecting fewer photons than were present. Dark counts, the other dominant source of imperfection in photodetection, could also be incorporated into the model, but this effect can be made very small with time-gating.

We specifically consider heralded SPDC states, using just one mode of the SPDC for the computation rather than both, to ensure that the state entering $U$ closely approximates Eq.~\ref{eq:input_state}. Without the heralding, the SPDC state is Gaussian, which is inconsistent with the boson-sampling model and not known to implement a classically hard algorithm \cite{bib:bartlett2003requirement, bib:gogolin2013boson}.

Combining Eqs.~\ref{eq:PDC_dist} \& \ref{eq:det_mn} we obtain the probability of detecting $t$ photons in the heralding arm of a single SPDC source,
\begin{eqnarray} \label{eq:pdc_det}
P^\mathrm{SPDC}_\mathrm{D}(t) &=& \sum_{i\geq t} P_\mathrm{D}(t|i) P^\mathrm{SPDC}(i) \nonumber \\
&=& \sum_{i\geq t} \binom{i}{t} \eta^t (1-\eta)^{i-t} P^\mathrm{SPDC}(i).
\end{eqnarray}
Thus the probability of detecting a single photon in the heralding arm is simply,
\begin{equation}
P^\mathrm{SPDC}_\mathrm{D}(1) = \sum_{i\geq 1} i \, \eta (1-\eta)^{i-1} P^\mathrm{SPDC}(i).
\end{equation}

\begin{figure}[!htb]
\includegraphics[width=0.8\columnwidth]{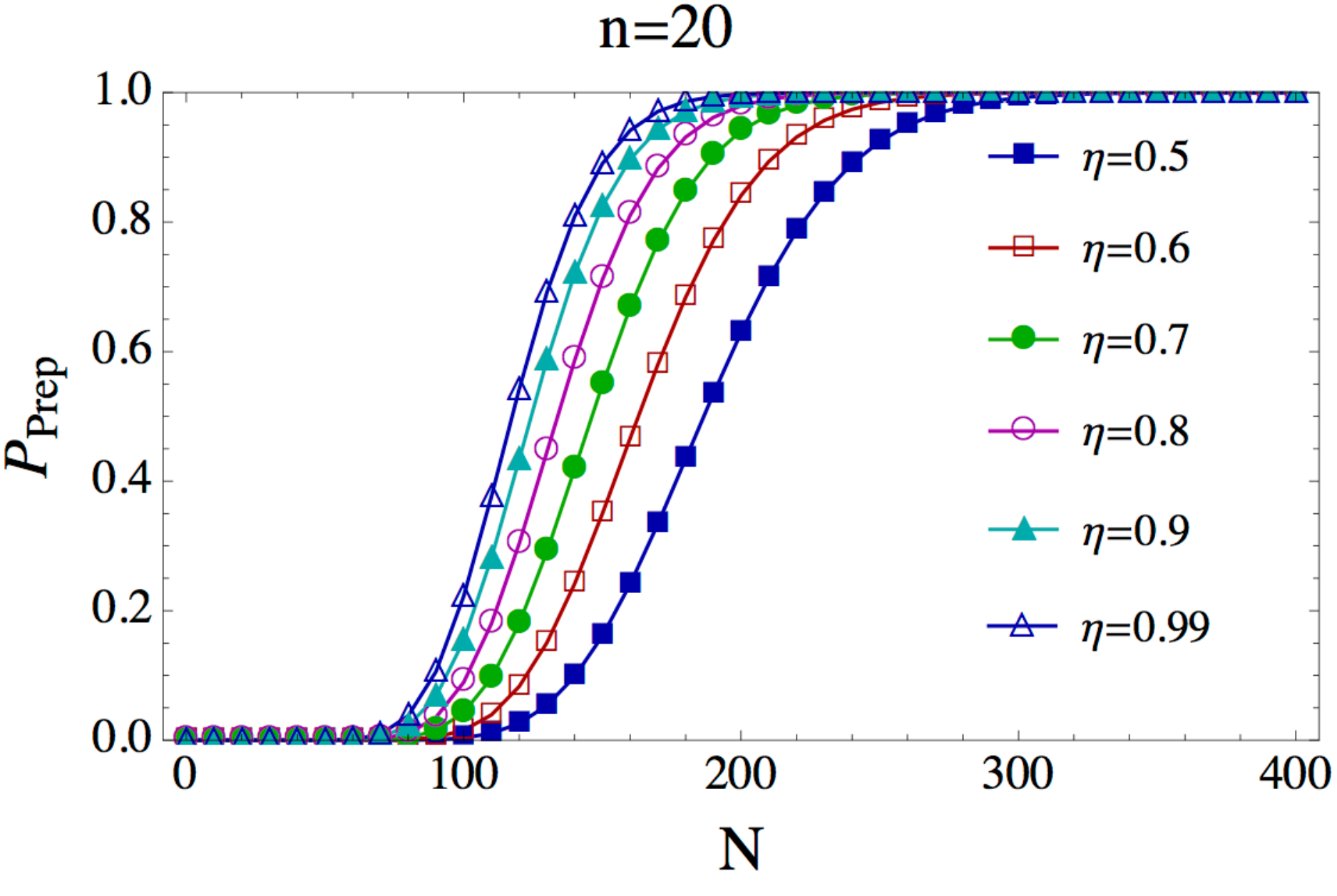}
\caption{Asymptotic behaviour of the state preparation success probability as a function of the number of SPDC sources, $N$, and detector efficiency, $\eta$, in the case where we are required to successfully herald \mbox{$n=20$} photons. In the limit of large $N$, $P_\mathrm{prep}$ approaches unity.} \label{fig:Pprep_asymptote}
\end{figure}

We will operate $N$ such heralded sources in parallel, where \mbox{$N\gg n$}. The probability that at least $n$ of the SPDC sources successfully herald is given by,
\begin{eqnarray} \label{eq:P_prep}
P_\mathrm{prep}(n) &=& \sum_{i \geq n} \binom{N}{i} [{P^\mathrm{SPDC}_\mathrm{D}(1)}]^{i} [1 - P^\mathrm{SPDC}_\mathrm{D}(1)]^{N-i}. \nonumber \\
\end{eqnarray}
In the limit of large $N$ this asymptotes to unity,
\begin{equation}
\lim_{N\to \infty}P_{\mathrm{prep}}(n)=1.
\end{equation}
The asymptotic behaviour of $P_{\mathrm{prep}}$ is illustrated in Fig.~\ref{fig:Pprep_asymptote}. Clearly, with a sufficiently large number of SPDC sources operating in parallel, we are guaranteed to successfully herald the required $n$ single photons.

Having successfully heralded at least $n$ SPDC sources, we employ a dynamic multiplexer \cite{bib:migdall2002tailoring} to route $n$ of the heralded states to the first $n$ modes of the boson-sampling interferometer $U$. We will assume the multiplexer is ideal in our analysis, although losses could be absorbed into the detector efficiency. Experimental progress has been made recently in developing active multiplexers \cite{bib:ma2011experimental, bib:LPOR201400027}.

Following the unitary network, number-resolving photodetection is applied. Because the photodetectors do not have unit efficiency we must post-select on events where all $n$ photons are detected. The post-selection probability scales as,
\begin{equation} \label{eq:exp_scaling}
P_\mathrm{post}(n) = \eta^n.
\end{equation}
Thus, the required detection efficiency exponentially asymptotes to unity for large $n$. This necessitates that future large-scale boson-sampling implementations will require extremely high efficiency photodetectors.

\begin{figure}[!htb]
\includegraphics[width=0.8\columnwidth]{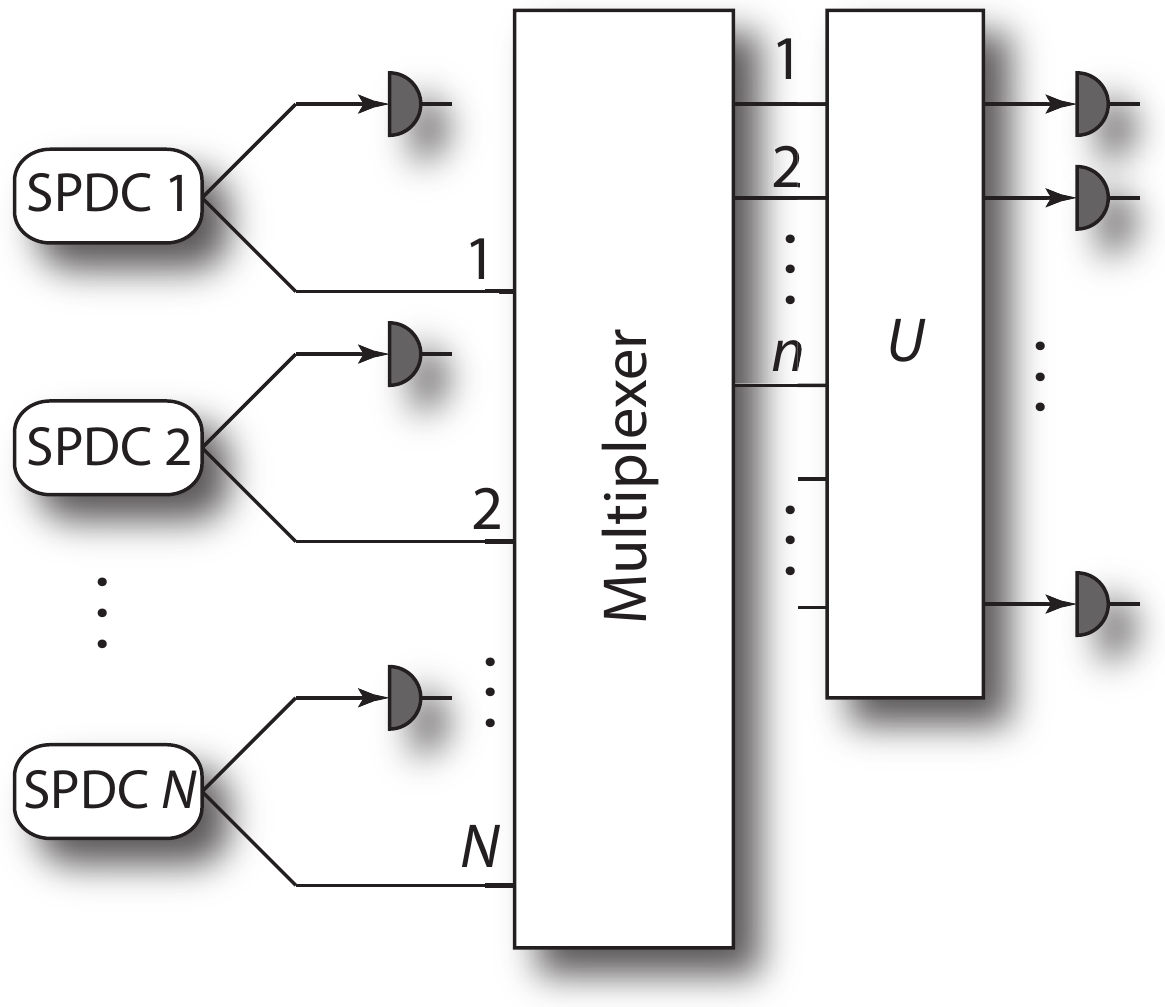}
\caption{Architecture for boson-sampling with SPDC sources. $N$ sources operate in parallel, each heralded by an inefficient single-photon number-resolving detection. It is assumed that \mbox{$N\gg n$}, which guarantees that at least $n$ photons will be heralded. The multiplexer dynamically routes the successfully heralded modes to the first $n$ modes of the unitary network $U$. Finally, photodetection is performed and the output is post-selected on the detection on all $n$ photons.} \label{fig:arch}
\end{figure}

The full architecture is illustrated in Fig.~\ref{fig:arch}. Note that the multiplexer is \emph{critical} to the operation of the device. Without the multiplexer we still have high likelihood of sampling from at least an $n$-photon input distribution. However, every time the device is run we are likely to sample from a different permutation of the vacuum and single photon states at the input, making it impossible to perform sampling on a consistent input. Thus, the multiplexing ensures that the input state is consistently of the form of Eq.~\ref{eq:input_state} if the photodetectors have perfect efficiency. The realistic case of inefficient photodetectors is presented next.

%
%

\section{Scalability of the architecture}

Having described a general architecture for boson-sampling based on SPDC sources, the pressing question is its scalability. The obvious scaling issue arises from Eq.~\ref{eq:exp_scaling}, whereby the photodetection efficiency must be exponentially close to unity. Unless error correction mechanisms are introduced, this scaling is inevitable and post-selection is the only avenue to guarantee successful operation of the device. However, no error correction has been described in the context of boson-sampling. Thus, we will focus on post-selected operation of the device, and address the question as to whether the device acts correctly in that context.

In the described architecture, the dominant error source is incorrect heralding of the SPDC states. In the limit of perfect detectors we are guaranteed to have prepared single-photon states. However, inefficient detectors introduce ambiguity in the heralding, creating a situation where higher-order photon number terms are perceived as single photon terms. For example, if a single photon is lost via detection inefficiency, the two photon state will be interpreted as a single photon state. This will corrupt the input state to the interferometer, yielding an input state different than Eq.~\ref{eq:input_state}.

For a single detector, the probability that we have prepared the $s$-photon Fock state, given that the detector has outcome $t$, is given by Bayes' rule,
\begin{eqnarray}
P_\mathrm{corr}(s|t) &=& \frac{P_\mathrm{D}(t|s) P^\mathrm{SPDC}(s)}{P^\mathrm{SPDC}_\mathrm{D}(t)} \nonumber \\
&=& \frac{\binom{s}{t} (1-\eta)^{s-t} \mathrm{tanh}^{2s} r}{\sum_{i\geq t} \binom{i}{t} (1-\eta)^{i-t} \mathrm{tanh}^{2i}r}.
\end{eqnarray}
We are interested in the case where we herald a single photon. Thus,  
\begin{equation}
P_\mathrm{corr}(1|1) = [1 - (1 - \eta)\,\mathrm{tanh}^{2}r]^2.
\end{equation}
$P_\mathrm{corr}(1|1)$ can be interpreted as the conditional probability that we have prepared the correct single photon state given that heralding was successful. For small pump powers (\mbox{$r\approx 0$}) the \emph{unconditional} probability of detecting a single photon approaches zero, although the \emph{conditional} probability approaches unity since there are negligible higher photon-number contributions.

The probability that a single photon is correctly heralded $n$ times in parallel, thereby preparing the $n$ copies of a single photon Fock state, is,
\begin{eqnarray} \label{eq:P_par}
P_\mathrm{par}(n) &=& [P_\mathrm{corr}(1|1)]^n \nonumber \\
&=& [1 - (1 - \eta)\,\mathrm{tanh}^{2}r]^{2n}.
\end{eqnarray}

We will require that, given $n$ heralded SPDC states, upon post-selection we correctly detect exactly $n$ photons the majority of the time. We will arbitrarily require that $P_\mathrm{post}(n) > \epsilon$, where $\epsilon$ is the lower bound on the probability that $n$ single photons are successfully detected at the output of the boson-sampling device. This puts a lower bound on the required photodetection efficiency of,
\begin{equation}
\eta = \sqrt[n]{\epsilon}.
\end{equation}

\begin{figure}[!htb]
\includegraphics[width=0.8\columnwidth]{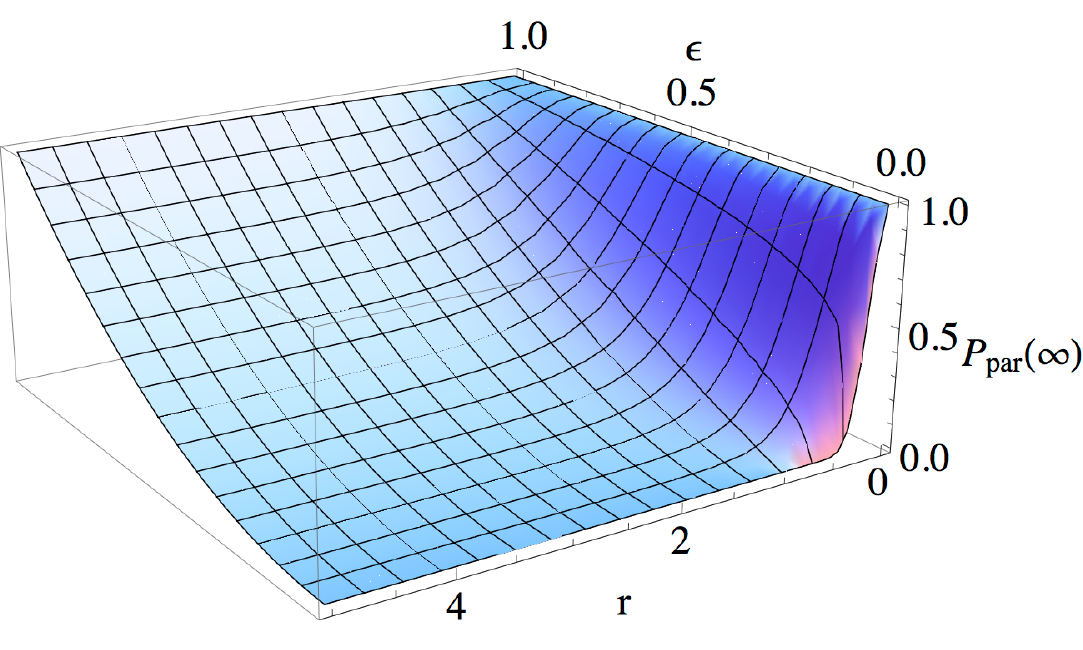}
\caption{(Color online) Probability that we are sampling from the correct input distribution in the limit of large $n$, obtained from Eq.~\ref{eq:Ppar}, plotted against the SPDC squeezing parameter $r$.} \label{fig:Ppar}
\end{figure}

Next we will assume that all photodetectors in the architecture have the same efficiency. Thus, we obtain that the probability of correctly preparing all $n$ photons via post-selected SPDC is,
\begin{equation}
P_\mathrm{par}(n) = [1 + (\sqrt[n]{\epsilon}-1)\,\mathrm{tanh}^{2}r]^{2n}.
\end{equation}
In the limit of large $n$ (i.e. large instances of boson-sampling), this asymptotes to,
\begin{equation} \label{eq:Ppar}
\lim_{n\to \infty} P_\mathrm{par}(n) = \epsilon^{2 \,\mathrm{tanh}^2 r}.
\end{equation}
For small $r$ this approaches unity, and in the limit of large $r$ to $\epsilon^2$. Thus, for $\epsilon = 1/2$, in the worst case scenario, we are sampling from the correct distribution in $1/4$ of the trials. This is shown in Fig.~\ref{fig:Ppar}.

Eq.~\ref{eq:Ppar} specifies the asymptotic probability of sampling from the correct input distribution, given that post-selection was successful. For small squeezing we sample from the correct input distribution most of the time, due to the lower probability of higher-order terms occurring. Thus, for experimentally realistic SPDC sources, provided that detector efficiencies are sufficiently high to enable post-selection, we have a high likelihood of correct boson-sampling and SPDC photon-number errors are negligible.

Conversely, we could require that $P_\mathrm{par} > \epsilon'$ from Eq. \ref{eq:P_par}, where $\epsilon'$ is the lower bound on the probability that a single photon is correctly heralded $n$ times in parallel before entering the multiplexer. Solving this for $\eta$ yields,
\begin{equation} \label{eq:eta2}
\eta = 1 + (\sqrt[2n]{\epsilon'} - 1)\,\mathrm{coth}^2 r.
\end{equation}
From Eq. \ref{eq:exp_scaling} we obtain an expression for the post-selection probability under the condition that we require a certain fidelity on the SPDC heralding,
\begin{equation} \label{eq:Ppost}
P_\mathrm{post}(n) = [1 + (\sqrt[2n]{\epsilon'} - 1)\,\mathrm{coth}^2 r]^n.
\end{equation}

\begin{figure}[!htb]
\includegraphics[width=0.8\columnwidth]{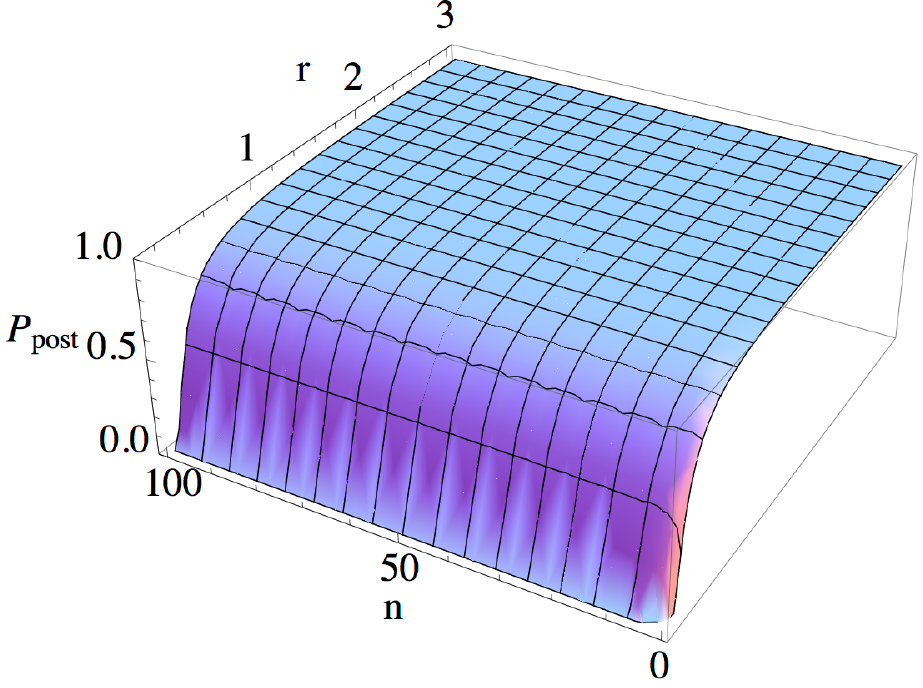}
\caption{(Color online) The post-selection probability $P_\mathrm{post}$ from Eq.~\ref{eq:Ppost} presented as a function of the squeezing parameter $r$ and $n$ single photons being correctly heralded in parallel before entering the multiplexer. Here we assume a fidelity of $\epsilon'=0.9$.} \label{fig:Ppost}
\end{figure}

Fig.~\ref{fig:Ppost} illustrates $P_\mathrm{post}(n)$ as a function of the squeezing parameter $r$ and the number of successfully routed photons $n$. We observe that for large $n$, post-selection is highly likely to succeed if the SPDC state preparation was successful to within error $\epsilon'=0.9$. We observe that in the limit of large $n$ and experimentally realistic values of \mbox{$r\approx1/2$}, boson-sampling using \mbox{$N\gg n$} SPDC sources is scalable. 

%
%

\section{Conclusion}

We presented a simple architecture for boson-sampling via multiplexed SPDC sources. We demonstrated that the SPDCs do not limit the scalability of the architecture. Rather, the single-photon detectors, whose efficiencies must increase exponentially with input photon number, limit the scalability. That is provided that detection efficiencies are sufficiently high to enable post-selected operation, the SPDCs will produce Fock states of sufficient fidelity to implement correct boson-sampling with high probability. Conversely, if detection efficiencies are sufficiently high to guarantee SPDC heralding with high fidelity, post-selection will succeed with high probability.

Thus, SPDC sources are a viable photon source technology for future large-scale demonstrations of boson-sampling, and experimentalists should prioritise improving detection efficiencies and developing single-photon multiplexing technologies. Additionally, existing SPDC sources will likely need significant improvement to increase squeezing purity and mode-matching.

While post-selection guarantees correct operation of a boson-sampling device, the required detection efficiencies scale unfavourably. Thus, future work should further address the question as to whether lossy boson-sampling is computationally hard \cite{bib:RohdeRalphErrBS}, as this could significantly reduce physical resource requirements. Other error models, such as mode-mismatch \cite{bib:RohdeLowFid12}, should also be further investigated.

The analysis presented could be applied to other post-selected linear optics protocols employing SPDCs as heralded Fock state sources.

%
%

\begin{acknowledgments}
This research was conducted by the Australian Research Council Centre of Excellence for Engineered Quantum Systems (Project number CE110001013). We thank S. E. Schm{\" a}mpling for helpful discussions and acknowledge the Air Force Office of Scientific Research and the National Science Foundation. 
\end{acknowledgments}

%
%

\bibliography{paper}

\end{document}